\documentclass[prl,twocolumn,showpacs,superscriptaddress,floatfix]{revtex4}

\usepackage{amsmath}
\usepackage{times}
\usepackage{txfonts}
\usepackage{graphicx}

\mathchardef\ogon="012C%
\newcommand{\as}{a\kern-0.22em\lower.40ex\hbox{$_{\ogon}$}}

\begin{document}

\title{Molecular production in two-component atomic Fermi gases}

\author{Jan Chwede\'nczuk}
\affiliation{Clarendon Laboratory, Department of Physics,  
  University of Oxford, Parks Road, Oxford, OX1 3PU, United Kingdom}
\affiliation{Physics Department, Warsaw University,
  Ho\.{z}a 69, PL-00-681 Warsaw, Poland}
\author{Krzysztof G{\'o}ral} 
\affiliation{Clarendon Laboratory, Department of Physics,  
  University of Oxford, Parks Road, Oxford, OX1 3PU, United Kingdom}
\affiliation{Center for Theoretical Physics, Polish Academy of 
  Sciences, Al.\ Lotnik\'ow 32/46, 02-668 Warsaw, Poland}
\author{Thorsten K\"{o}hler}
\affiliation{Clarendon Laboratory, Department of Physics,  
  University of Oxford, Parks Road, Oxford, OX1 3PU, United Kingdom}
\author{Paul S.\ Julienne}
\affiliation{Atomic Physics Division, National Institute of Standards and 
  Technology, 100 Bureau Drive Stop 8423, Gaithersburg, Maryland
  20899-8423, USA}

\begin{abstract}
We provide a practical approach to the molecular production via linear 
downward sweeps of Feshbach resonances in degenerate Fermi gases containing 
incoherent mixtures of two atomic spin states. We show that the efficiency of 
the association of atoms is determined just by the Landau-Zener parameter in 
addition to the density of the gas. Our approach of pairwise summation of the 
microscopic binary transition probabilities leads to an intuitive explanation 
for the observed
%50\%
saturation
%limit
of the molecular production and 
recovers all atomic loss curves of C.A.~Regal {\em et al.} 
[Nature (London) \textbf{427}, 47 (2003)] as well as K.E.~Strecker {\em et al.}
[Phys.~Rev.~Lett.~\textbf{91}, 080406 (2003)] without adjustable parameters.  
\end{abstract}
\date{\today}
\maketitle

Ever since the achievement of Bose-Einstein condensation in a dilute 
vapour of $^{87}$Rb atoms, there has been considerable interest in extending 
the range of species for studies of cold gases. Several experiments 
\cite{Donley02,Claussen03,Regal03,Herbig03,Duerr03,Strecker03,Cubizolles03,JochimPRL03,Xu03} 
have now demonstrated the production of cold molecules via 
magnetic field tunable Feshbach resonances. One of the techniques relies upon 
the adiabatic association of pairs of cold atoms to molecules through a linear 
sweep of a Feshbach resonance level across the zero energy threshold of the 
colliding atoms. This technique has been applied to atomic Bose-Einstein 
condensates \cite{Herbig03,Duerr03,Xu03}, as well as degenerate two-component 
Fermi gases \cite{Regal03,Strecker03} containing a balanced incoherent mixture 
of Zeeman levels of a single atomic species. In contrast to cold Bosons, 
degenerate Fermions inevitably occupy a range of single particle energy levels 
up to the Fermi edge. There have been concerns that the spread in collision 
momenta of Fermionic pairs could prevent an efficient molecular production in 
an adiabatic sweep. Despite this intrinsic difficulty, a significant molecular 
conversion of up to approximately
%of up to a limit of
50\% of the atoms was reported in 
Refs.~\cite{Regal03,Strecker03}, and higher
conversion efficiencies have been reported when much
slower ramp speeds are used \cite{Cubizolles03,Greiner03}.
%.
The few attempts
\cite{Javanainen04,Pazy04,Tikhonov04,Williams04} to explain these observations are all 
based on rather involved field theoretic models. 
While Refs.~\cite{Pazy04,Tikhonov04} provide simulation methods for the 
molecular production in the presence of Fermi seas, Ref.~\cite{Javanainen04} 
suggests the maximum molecular production efficiency of 50\% in 
Ref.~\cite{Regal03} to be a consequence of a significant temperature 
dependence of the conversion process. 

In this letter we derive a practical universal formula for the 
molecular production via a linear sweep of the magnetic field strength $B$ in 
a degenerate two-component Fermi gas in the short time limit where only a 
single collision of a given atom occurs during the sweep. Our analytic 
treatment relies upon a
pairwise summation of the probabilities for the atomic association, which
naturally explains the fast ramp observations
%50\% limit
of Refs.~\cite{Regal03,Strecker03}. 
We show that the molecular production is largely insensitive to the cold 
collision momenta and depends only on the product 
$a_\mathrm{bg} \Delta B$ of the background scattering length $a_\mathrm{bg}$ 
and the resonance width $\Delta B$, in addition to the atomic mass $m$, the 
ramp speed $\dot{B}$ as well as the density of the gas. Despite its 
simplicity, our analytic formula leads to predictions for the molecular 
production that are virtually indistinguishable from the observations of 
Refs.~\cite{Regal03,Strecker03} within the experimental error bars.
 
We shall first discuss the association of atomic pairs in the 
absence of the surrounding gas. A homogeneous magnetic field 
$B$ splits the atomic hyper-fine angular momentum states into 
a set of Zeeman levels $(f,m_f)$, labelled by the projection quantum 
number $m_f$ of the total atomic spin along the axis of the magnetic field 
and the $f$ value with which it correlates adiabatically at zero field.
Indistinguishable Fermions with equal spins are not subject to $s$-wave 
scattering and do therefore not contribute to molecular formation. 
We shall thus consider pairs of atoms whose 
internal states correspond to different Zeeman levels at asymptotically large 
inter-atomic separations. We denote the $s$-wave binary scattering channel 
of such a pair of unlike Fermions as the entrance channel. 

\begin{figure}[htb]
\begin{center}
\includegraphics[width=\columnwidth,clip]{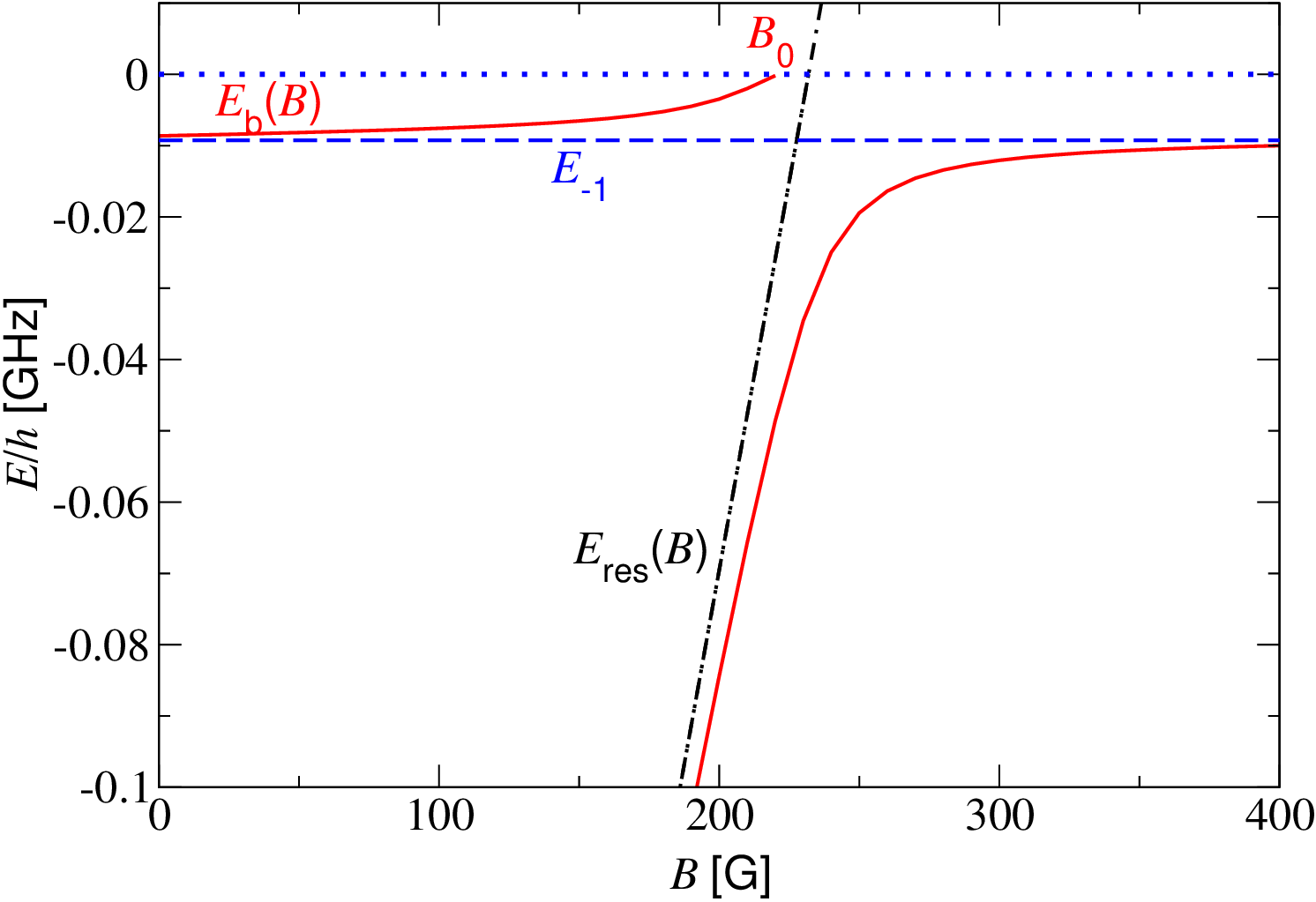}
\caption{The binding energy $E_\mathrm{b}(B)$ (solid curve) of the 
  highest excited dressed vibrational bound state of a pair of $^{40}$K atoms 
  versus the magnetic field strength $B$ in the vicinity of the 224 G zero 
  energy resonance \cite{Regal03}. The zero of energy (dotted line) is chosen, 
  for each magnetic field strength, as the Zeeman energy of a pair of 
  distinguishable Fermions at asymptotically large spatial separation. The 
  zero of the dressed state energy $E_\mathrm{b}$ coincides with the 
  singularity of the scattering length in the entrance channel at the magnetic 
  field strength $B_0$. The dotted dashed line indicates the energy 
  $E_\mathrm{res}(B)$ of the bare Feshbach resonance state. The dashed line 
  corresponds to the energy $E_{-1}$ of the highest excited bare vibrational 
  level of the entrance channel potential.}
\label{fig:Eb40K}
\end{center}
\end{figure}

The physical concept of associating a pair of cold atoms is closely related to 
the variation of its energy spectrum under adiabatic changes of $B$. 
Figure \ref{fig:Eb40K} shows a coupled 
channels calculation of these energies for the example of the interacting 
pairs of $^{40}$K atoms in the $(f=9/2,m_f=-9/2)$ and $(f=9/2,m_f=-5/2)$ 
Zeeman states of Ref.~\cite{Regal03}. The figure shows that when $B$ 
varies from high to low fields across the zero energy resonance at 
$B_0=224$ G (1 G$=10^{-4}$ T), a molecular bound state 
$\phi_\mathrm{b}(B)$ appears whose energy $E_\mathrm{b}(B)$ is 
indicated by the solid curve. The zero of the binding energy 
$E_\mathrm{b}(B)$ coincides with the magnetic field strength $B_0$ 
at which the scattering length has a singularity. 

The strong magnetic field dependence of the cold inter-atomic 
collisions is due to the near degeneracy of the energy $E_\mathrm{res}(B)$ 
(dotted dashed line in Fig.~\ref{fig:Eb40K}) of a single closed channel 
vibrational state $\phi_\mathrm{res}$ (the Feshbach resonance level) with the 
threshold for dissociation in the entrance channel (dotted line in 
Fig.~\ref{fig:Eb40K}). This allows us to use a two-channel configuration 
interaction representation \cite{Mies00,GKGTJ03} in terms of the bare basis 
states, i.e.~the states in the absence of inter-channel coupling. These bare 
states include the Feshbach resonance level $\phi_\mathrm{res}$ and the 
entrance channel continuum states $\phi_\mathbf{p}^{(+)}$ associated with the 
collision momentum $\mathbf{p}$. The kinetic energy $p^2/m$ 
as well as the magnetic moment $\mu_\mathrm{res}=dE_\mathrm{res}/dB$ of the 
resonance level are independent of $B$. Consequently, the determination of the 
probability for the association of an atom pair with the relative momentum 
$\mathbf{p}$ can be described by the Landau-Zener approach 
\cite{Mies00,GKGTJ03}. We note, however, that although we have chosen a bare 
state representation, the bound molecules produced in a linear sweep of the 
magnetic field strength are described by the dressed state 
$\phi_\mathrm{b}(B)$, which accounts for the inter-channel coupling. 

The relative momenta in a zero temperature two-component Fermi gas 
are limited by the maximum of the Fermi momenta of each component, whose 
associated energy scales are typically on the order of $\mu$K (in units of 
the Boltzmann constant). These energies are too small to be resolved in 
Fig.~\ref{fig:Eb40K}. In the limited range of cold collision momenta
the probability $P$ for the association of a 
single pair of  Fermions in a large sample volume $\mathcal{V}$ is  
isotropic, i.e.~it depends just on the wave number $k=p/\hbar$ associated 
with the collision momentum $\mathbf{p}$. Neglecting 
phenomena related to the presence of the surrounding 
gas, considerations described in Ref.~\cite{GKGTJ03} lead 
to the formula \cite{multiplecrossings}: 
\begin{equation}
P(k)=-\frac{(2\pi)^3}{\mathcal{V}}
\frac{1}{4\pi k^2}\frac{\partial}{\partial k}\exp
\left(
-\frac{\mathcal{V}\delta_\mathrm{LZ}}{3\pi}k^3
\right).
\label{multipleLZ}
\end{equation}
The inter-atomic interactions enter the probability in terms of the 
Landau-Zener parameter
\cite{Mies00,GKGTJ03}:
\begin{equation}
  \delta_\mathrm{LZ}=\frac{(2\pi\hbar)^3
    \left|\langle\phi_\mathrm{res}|W|\phi_\mathbf{p}^{(+)}\rangle\right|^2}
	{\mathcal{V}|\dot{B} \mu_\mathrm{res}|}
	  =\frac{4\pi\hbar |a_\mathrm{bg}\Delta B|}
	  {\mathcal{V}m|\dot{B}|}.
\end{equation}
Here $W(r)$ is the potential describing the inter-channel coupling. We note
that $\langle\phi_\mathrm{res}|W|\phi_\mathbf{p}^{(+)}\rangle$ and 
consequently $\delta_\mathrm{LZ}$ are independent of $k$ provided 
that $|a_\mathrm{bg}|$ is sufficiently small to satisfy the relation 
$k|a_\mathrm{bg}|\ll 1$. The $k$ dependence of Eq.~(\ref{multipleLZ}) is due 
to downward transitions into lower energetic continuum states that compete 
with the molecular formation in an isolated two-body system \cite{Mies00}.  
 
We shall show in the following that the molecular production in the limit 
of fast magnetic field ramps as well as in the opposite saturation limit can 
be determined in a straightforward manner, provided that multiple 
collisions of a given atom can be neglected. To this end, we shall 
first consider the limit of ramp speeds sufficiently high that 
Eq.~(\ref{multipleLZ}) is well approximated by first order perturbation theory 
in the inter-channel coupling. A linear expansion 
of Eq.~(\ref{multipleLZ}) in $\delta_\mathrm{LZ}$ then shows
that $P(k)$ is independent of $k$, and reduces to the linearised Landau-Zener 
formula for a single curve crossing \cite{Mies00,GKGTJ03}:
\begin{equation}
 P= 2\pi\delta_\mathrm{LZ}.
  \label{LZ}
\end{equation}
Despite the fact that the association of a given atom pair in the 
large sample volume $\mathcal{V}$ is a rare event, the molecular production 
in an incoherent mixture of $N$ Fermions can be rather efficient because 
each of the $N_1$ atoms in one of the spin states has all $N_2$ 
atoms of the other component to interact with. In accordance with classical 
probability theory, the total number $N_\mathrm{b}$ of diatomic 
molecules in the state $\phi_\mathrm{b}$ is given by the pairwise incoherent 
sum of the microscopic transition probabilities: 
\begin{equation}
  N_\mathrm{b}=
  N_1N_2 P.
  \label{probabilistic}
\end{equation} 
Here $N_1N_2$ is the number of pairs of unlike atoms. 
Equations (\ref{LZ}) and (\ref{probabilistic}) 
then lead to the following fraction of atoms associated to molecules:
\begin{equation}
  2N_\mathrm{b}/N= 2\alpha(1-\alpha)N P
  = 2\alpha(1-\alpha)\left[2\pi n\mathcal{V} \delta_\mathrm{LZ}\right].
  \label{probabilistic2}
\end{equation}  
Here $n=N/\mathcal{V}$ is the local density and $\alpha$ is the minimum of 
the fractions $N_1/N$ and $N_2/N$. Since $\mathcal{V}\delta_\mathrm{LZ}$ is 
independent of $\mathcal{V}$, Eq.~(\ref{probabilistic2}) depends just on the 
local density of the gas. The pre-factor $\alpha(1-\alpha)$ accounts for the 
size of the sub-ensemble of pairs of unlike Fermions and the
factor of two indicates the number of atoms per molecule, while 
$2\pi n\mathcal{V} \delta_\mathrm{LZ}$ can be interpreted as a density 
dependent Landau-Zener transition probability \cite{Mies00,GKGTJ03}. 

The pairwise summation of microscopic transition probabilities is justified 
when the number of molecules produced is small in comparison to the total 
number of atoms \cite{GKGTJ03,KGTKKB04}. In this limit of high ramp speeds 
the molecular production does not crucially depend on the presence of the 
Fermi seas. As the ramp speed decreases, Eq.~(\ref{multipleLZ}) becomes 
momentum dependent, since, unlike for fast ramps, there can be decay 
during the ramp to lower energy continuum states of the two-body system 
with $E>0$.  In the presence of the surrounding two-component 
Fermi gas, however, these downward transitions would lead into 
occupied modes and are therefore prevented. In order to extend 
Eq.~(\ref{probabilistic2})  
to lower ramp speeds, we shall thus follow the approach of 
Refs.~\cite{Mies00,GKGTJ03} and replace the density dependent linearised
Landau-Zener formula $2\pi n\mathcal{V} \delta_\mathrm{LZ}$ by the 
complete Landau-Zener transition probability 
$1-\exp(-2\pi n\mathcal{V} \delta_\mathrm{LZ})$ for a linear curve 
crossing in the absence of downward transitions to other continuum states. 
The corrections to this procedure involve the dynamic depletion of the 
available pairs during the atomic association. Comparisons between 
genuinely many-body approaches and the Landau-Zener formula 
\cite{GKGTJ03} for the case of a condensate of identical Bosons
show that the dynamic depletion of the pairs changes 
the functional dependence of the  molecular production efficiency on 
the ramp speed. The absolute magnitude of these 
corrections, however, is typically small in comparison with the total 
depletion of the atoms and can be neglected.
The molecular production efficiency in a two-component Fermi gas may thus 
be approximated by:
\begin{equation}
  \frac{2N_\mathrm{b}}{N}=2\alpha(1-\alpha)
  \left[1-\exp\left(-n\frac{8\pi^2\hbar |a_\mathrm{bg}\Delta
      B|}{m|\dot{B}|}\right)\right].
  \label{efficiency}  
\end{equation}

%Equation (\ref{efficiency}) shows that the
The molecular production efficiency can not exceed 50\%, given the
ansatz in Eq. (\ref{efficiency}). This limit 
is quite intuitive as, given only one chance to associate, a Fermion in the 
larger one of the two components can find an interacting partner with a 
probability of 1/2 at best. The saturated efficiency in mixtures 
with unbalanced populations of the spin components can only be lower and 
is given by $2\alpha(1-\alpha)$. The exponential function in 
Eq.~(\ref{efficiency}) provides a smooth interpolation between the 
straightforward asymptotic limits of the ramp speed.

The range of validity of our approach is set by the assumption that only a 
single collision of a given atom can occur during the association process. 
When the ramp speed is chosen sufficiently low for multiple collisions to
become significant, efficiencies exceeding 50\% can result
% the 50\% limit may be exceeded 
\cite{Cubizolles03,Greiner03}
%by
from the same physical
mechanisms that can convert an atomic gas to a molecular gas at constant
magnetic field strength \cite{JochimPRL03,JochimScience03,Zwierlein03}.
To this end, the experiments of Ref.~\cite{Greiner03} applied magnetic field 
ramps at least ten times slower than the slowest ramp of Ref.~\cite{Regal03} 
for Feshbach resonance parameters and densities comparable to those of 
Ref.~\cite{Regal03}. 

\begin{figure}[htb]
  \begin{center}
    \includegraphics[width=\columnwidth,clip]{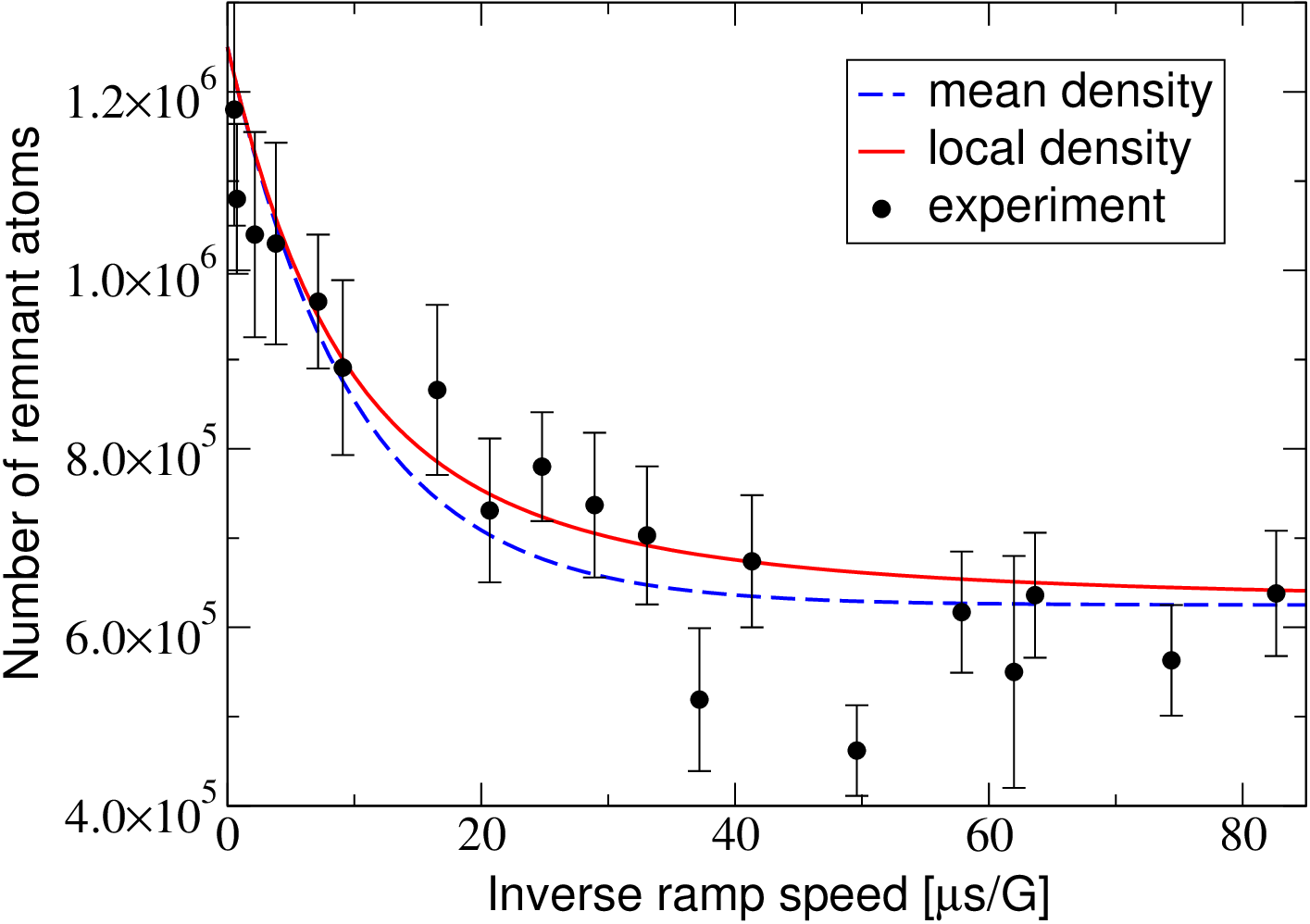}
    \caption{The number of atoms remaining versus the inverse ramp speed 
      $1/\dot{B}$ for a linear sweep of the magnetic field strength across the 
      224 G zero energy resonance in a two-component Fermi gas of $^{40}$K. 
      The experimental data of Ref.~\cite{Regal03} (circles) are compared to 
      the results of the pairwise summation approach of 
      Eq.~(\ref{efficiency}) with $\alpha=1/2$. 
      The solid and dashed curves indicate local and mean density 
      approximations, respectively.
      }
    \label{fig:40K}
  \end{center}
\end{figure}

To compare the predictions of Eq.~(\ref{efficiency}) to the
experimental results of Refs.~\cite{Regal03,Strecker03}, we have applied the 
local density approximation, i.e.~we have averaged Eq.~(\ref{efficiency}) in 
accordance with the density distribution of the trapped gases. We have 
determined the density profiles from the zero temperature Thomas-Fermi 
approximation \cite{Butts97}, adapted to the experimental trap geometries. 
Figure \ref{fig:40K} shows the number $N-2N_\mathrm{b}$ of free $^{40}$K 
atoms remaining in the 
experiments of Ref.~\cite{Regal03} 
as compared to the predictions 
of Eq.~(\ref{efficiency}). We have estimated the total number of atoms to be 
$N=1.25\times 10^6$ (cf.~Fig.~\ref{fig:40K}) and used the background 
scattering length $a_\mathrm{bg}=174\, a_0$ \cite{Regal03-2} 
($a_0=0.052918$ nm), the resonance width $\Delta B=9.7$ G 
\cite{Regal03-2}, and the radial angular trap frequency 
$\omega_\mathrm{r}=2\pi\times 215$ s$^{-1}$, which, together with the aspect ratio 
of 70 \cite{Regal03}, largely recover the experimental peak density of 
$n_\mathrm{pk}=1.4\times10^{13}$ cm$^{-3}$ \cite{Regal03}. This procedure 
yields a mean 
density of $n=9\times10^{12}$ cm$^{-3}$. The predictions of the mean 
and local density approaches to Eq.~(\ref{efficiency}) clearly follow the 
slope as well as the magnitude of the experimental data.
We obtain a similar agreement with unpublished lower density  
($n_\mathrm{pk}=9\times10^{12}$ cm$^{-3}$) data of Regal {\em et al.} 
\cite{Regalprivate}.

\begin{figure}[htb]
  \begin{center}
    \includegraphics[width=\columnwidth,clip]{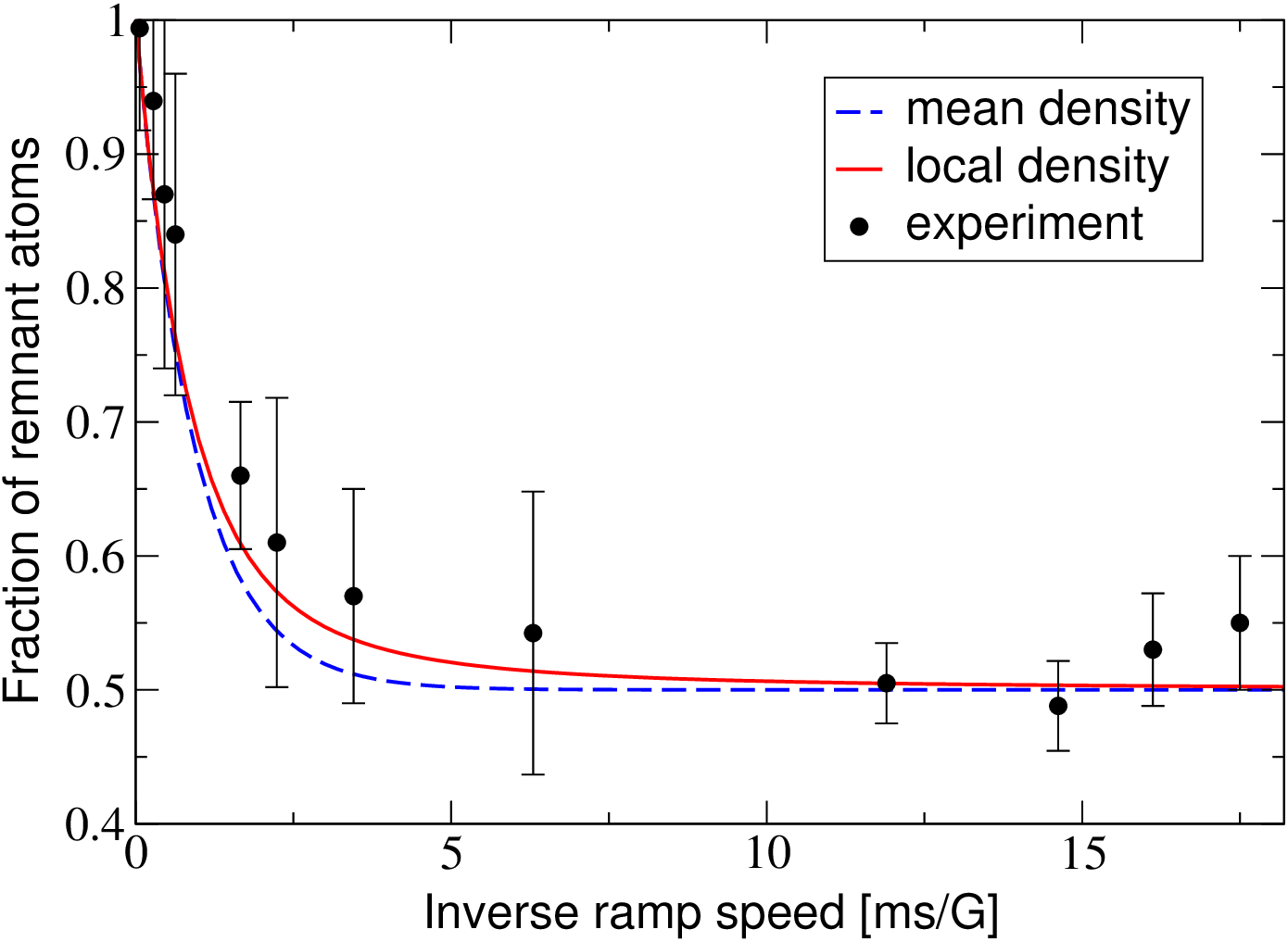}
    \caption{The fraction $1-2N_\mathrm{b}/N$ of remnant atoms versus 
      the inverse ramp speed $1/\dot{B}$
      for a linear sweep of the magnetic field strength across
      the 543 G zero energy resonance in a two-component Fermi gas of 
      $^6$Li. The circles indicate the experimental data of 
      Ref.~\cite{Strecker03}. The solid and dashed curves are predictions of
      Eq.~(\ref{efficiency}) with $\alpha=1/2$ in the local and mean 
density approximations, respectively.
      }
    \label{fig:6Li}
  \end{center}
\end{figure}

To further confirm the validity of Eq.~(\ref{efficiency}), Fig.~\ref{fig:6Li} 
shows an analogous comparison with the results of
Ref.~\cite{Strecker03}. In these experiments the molecules were produced
by a linear sweep of the magnetic field strength across the 543 G zero 
energy resonance of $^6$Li in the entrance channel associated with the 
set of quantum numbers $(f=1/2,m_f=+1/2)$ and $(f=1/2,m_f=-1/2)$ of the 
interacting atoms. We have performed an exact coupled channels calculation 
to determine the resonance parameters. From these considerations we obtain 
$|a_\mathrm{bg}\Delta B|=5.916\, a_0$G. The Thomas-Fermi 
density profile used in the local density approach is determined by the Fermi 
energy $E_\mathrm{F}$, which in the case of the combined harmonic and 
box-like trapping potential of Ref.~\cite{Strecker03} is given by 
$E_\mathrm{F}=\left[15 \pi N  \hbar^3 
\omega_\mathrm{r}^2/(8\sqrt{2m}L)\right]^{2/5}$. 
Here $\omega_\mathrm{r}=2\pi\times 800$ s$^{-1}$
is the angular frequency of the radial harmonic trap and $L=480\ \mu$m 
is the size of the axial box-like potential \cite{Strecker03}. This 
leads to a mean density of $n=4\times 10^{12}$ cm$^{-3}$.

The agreement with the experimental data in Figs.~\ref{fig:40K} and 
\ref{fig:6Li} suggests that the pairwise summation approach recovers all the 
relevant aspects of the physics involved. This implies that the molecular 
production efficiency is determined by the density of the gas, the ramp speed 
and the atomic mass, in addition to a single universal parameter of the 
resonance enhanced collision physics, the product of the background scattering 
length and the resonance width. The density profile provides a direct 
dependence of the molecular production on the Fermi statistics. Our results 
indicate, however, that even a mean density approximation yields a rather 
useful estimate of the number of molecules produced. Our ansatz
provides an excellent approximation for the
molecular production efficiency in the limit of very
fast downward sweeps of the Feshbach
resonance level in two-component Fermi gases
and gives a 50\% efficiency as a natural limit, provided
that phenomena related to multiple collisions of a given
atom are negligible.
%Our approach shows that the observed maximum 50\% molecular production
%efficiency provides a natural limit for the technique of adiabatic downward sweeps of the
%Feshbach resonance level in two component Fermi gases, provided that 
%phenomena related to multiple collisions of a given atom are negligible.
The pairwise summation approach has a rather wide range of 
applicability in dilute gases, including cold Bosons 
\cite{Mies00,GKGTJ03,KGTKKB04} and Fermions as well as their mixtures. 
The main requirement is the diluteness of the gas in combination with a 
binary transition, which is sufficiently fast that multiple collisions
can be neglected. Such a binary transition may be realised by a single linear 
sweep of the magnetic field strength, but it can as well involve fast sequences
of magnetic field pulses \cite{KGTKKB04} applied in the experiments
of Refs.~\cite{Donley02,Claussen03}.

We are particularly grateful to Cindy Regal and Randy Hulet for providing 
their experimental data to us. This research has been supported by the 
ESF Programme BEC2000+ and the Polish Ministry of Scientific Research and 
Information Technology under grant PBZ-MIN-008/P03/2003 (J.Ch.), the Royal 
Society (K.G. and T.K.) and the US Office of Naval Research (P.S.J.).

\end{document}